\documentclass[conference]{IEEEtran}
\IEEEoverridecommandlockouts
\usepackage[braket, qm]{qcircuit}
\usepackage{graphicx}
\usepackage{physics}
\usepackage{cite}
\usepackage{amsmath,amssymb,amsfonts}
\usepackage{graphicx}
\usepackage{textcomp}

\usepackage[ruled,vlined,linesnumbered]{algorithm2e}
\SetKwInOut{Input}{Input}
\SetKwInOut{Output}{Output}
\SetKwComment{Comment}{$\triangleright$\ }{}
\usepackage{xcolor}
\usepackage{subcaption}
\usepackage{amsmath}
 \usepackage[hidelinks]{hyperref}
\usepackage{pgffor}
\usepackage{adjustbox}

\usepackage{threeparttable}

\usepackage[numbers,compress]{natbib}

\definecolor{quantumweek-1}{RGB}{182,214,186}
\definecolor{quantumweek-2}{RGB}{214,205,182}
\definecolor{quantumweek-3}{RGB}{182,190,214}
\definecolor{quantumweek-4}{RGB}{214,182,192}
\definecolor{quantumweek-5}{RGB}{81,93,130}
\definecolor{quantumweek-6}{RGB}{39,87,45}

\def\BibTeX{{\rm B\kern-.05em{\sc i\kern-.025em b}\kern-.08em
    T\kern-.1667em\lower.7ex\hbox{E}\kern-.125emX}}
\begin{document}

\title{Quantifying Teleportation Overhead in Distributed Unitary
Coupled-Cluster Ans\"{a}tze\\
\thanks{We would like to acknowledge the Government of Canada’s New Frontiers in Research Fund (NFRF), for grant NFRFE-2022-00226, and the Quantum Software Consortium (QSC), financed under grant \#ALLRP587590-23 from the Natural Sciences and Engineering Research Council of Canada (NSERC) Alliance Consortia Quantum Grants.
H. T. S. and Z. W. acknowledge Mitacs Canada for supporting their research.
The authors acknowledge the use of Anthropic's Claude~\cite{claude2026} and Google's Gemini~\cite{gemini2026} to draft
portions of this manuscript, which were reviewed and edited
by the authors before submission.
}
}

\author{
\IEEEauthorblockN{
Grier M. Jones\IEEEauthorrefmark{1}\IEEEauthorrefmark{3}\IEEEauthorrefmark{4},
Hassan Tariq Shafi\IEEEauthorrefmark{2},
Zixuan Wang\IEEEauthorrefmark{2},
Thomas Trenty\IEEEauthorrefmark{2},
Zachary Vernec\IEEEauthorrefmark{2},
Hans-Arno Jacobsen\IEEEauthorrefmark{2}\IEEEauthorrefmark{3}\IEEEauthorrefmark{5}
}

\IEEEauthorblockA{
\IEEEauthorrefmark{1}Department of Chemical and Physical Sciences\\
\IEEEauthorrefmark{2}Department of Computer Science\\
\IEEEauthorrefmark{3}The Edward S. Rogers Sr. Department of Electrical and Computer Engineering\\
University of Toronto\\
Toronto, Ontario, Canada\\
\IEEEauthorrefmark{4}grier.jones@utoronto.ca
\IEEEauthorrefmark{5}jacobsen@eecg.toronto.edu
}

}

\IEEEoverridecommandlockouts 

\maketitle

\begin{abstract}
Distributed quantum computing (DQC) has been proposed as a way to scale quantum algorithms for practical applications beyond monolithic quantum processor architectures.
Among these applications, quantum chemistry is widely regarded as one of the most promising use cases for quantum computing.
In this work, we estimate the distributed-resource requirements of unitary coupled-cluster (UCC) ans\"{a}tze for quantum chemistry, focusing on unitary coupled-cluster singles and doubles (UCCSD), unitary pair coupled-cluster doubles (UpCCD), and unitary pair coupled-cluster with generalized singles and doubles (UpCCGSD) circuits for hydrogen chains.
We focus on a teleportation-based approach to DQC, quantitatively comparing a naive distribution method to the output of the TeleSABRE~\cite{russo2025telesabre} algorithm. 
For both approaches, we estimate the cost of handling nonlocal two-qubit gates across a fixed midpoint or quarter-point partition, reporting Bell-pair/classical-communication costs in teleportation.
Across Jordan-Wigner and Bravyi-Kitaev, we find that UpCCD with spin-blocked Jordan-Wigner ordering gives the most favorable scaling, while UCCSD incurs substantially larger distributed-resource requirements.
\end{abstract}

\begin{IEEEkeywords}
Electronic Structure Theory, Distributed Quantum Computing, Quantum Chemistry, Gate Teleportation 
\end{IEEEkeywords}

\section{Introduction}

Quantum chemistry is a promising near-term application for quantum computing~\cite{reiher2017elucidating}.
A common task in electronic structure calculations is finding the ground-state energy using the Schrödinger equation with a nonrelativistic Hamiltonian, invoking the Born–Oppenheimer approximation in the absence of an external field. 
In principle, the ground-state energy can be determined exactly using full configuration interaction (FCI) within a finite basis set. 
In practice, FCI is computationally intractable on classical hardware for systems where the active space---defined by $N_{e}$ electrons in $N_{o}$ spatial orbitals and denoted as $(N_{e}, N_{o})$---extends beyond $(22, 22)$~\cite{vogiatzis_pushing_2017} or~$(26, 23)$~\cite{gao_distributed_2024}.
Due to the exponential scaling of the FCI expansion, approximate methods, such as ans\"{a}tze based on UCC theory, have been introduced to solve the electronic structure problem on quantum computers~\cite{anand2022quantum,cao_quantum_2019}.
While algorithms like quantum phase estimation (QPE)~\cite{aspuru-guzik_simulated_2005,abrams_simulation_1997,abrams_quantum_1999,lanyon_towards_2010,whitfield_simulation_2011,aspuru-guzik_photonic_2012} and variational quantum eigensolver (VQE)~\cite{peruzzo_variational_2014,cerezo_variational_2021,mcclean_theory_2016,bharti_noisy_2022} can compute molecular properties, they remain constrained by qubit count and quality. 
To bypass these hardware limits, techniques like entanglement forging~\cite{eddins_doubling_2022} and distributed quantum computing (DQC) are actively used.
\ignorespaces
\\
\indent In our previous work~\cite{jones2025analyzing}, we analyzed several common quantum chemistry ans\"{a}tze in the context of circuit cutting, finding in particular that most UCC-derived ans\"{a}tze have a prohibitive sampling overhead.
In this work, we instead consider explicit resource estimations for teleportation-based distribution.
Specifically, we evaluate and contrast the estimated Bell pairs of an unoptimized (naive) circuit partition with those of TeleSABRE~\cite{russo2025telesabre}, a method for quantum layout and routing on multi-core, teleport-based quantum processing units (QPUs).
We perform resource analysis for implementing UpCCD, UpCCGSD, and UCCSD ans\"{a}tze on two coupled 120-qubit square-lattice QPUs, similar to IBM Quantum's 120-qubit Nighthawk. 
Our results highlight that naive partitioning has similar results for both fermion-to-qubit mappings, with the notable exception of UCCSD, and that the scaling is similar when partitioning into halves and quarters. 
We further note that TeleSABRE greatly decreases the number of Bell pairs required compared to naive layouts without routing.
\ignorespaces
\\
\indent This paper is organized as follows. In Section~\ref{section:theory}, we give background information required for our work: our application field (quantum chemistry), our primitive operations for distributed quantum computing (gate and state teleportation), and our choice of distributed resource optimization algorithm (TeleSABRE).
In Section~\ref{section:compworkflow}, we describe the computational workflow under consideration, including in particular our choice of circuits and our partitioning approach. 
In Section~\ref{section:resultsanddiscussion}, we present our partitioning results and discussion, and in Section~\ref{section:conclusions} we present our conclusions.
\section{Theoretical Framework}\label{section:theory}
\subsection{Quantum Chemistry}\label{subsection:EStheory}
One of the key goals of quantum chemistry is to understand chemical phenomena by solving the Schr\"{o}dinger equation, $\hat{H}\Psi=E\Psi$, where $\Psi$ is a wave function describing the state of the system, $\hat{H}$ is the molecular Hamiltonian, and $E$ is the energy of the system, which is an eigenvalue of the Hamiltonian.~\cite{helgaker2013molecular}.
While Hartree-Fock (HF) is the starting point for many electronic structure theory calculations, the resulting HF wave function fails to capture the energy associated with electron correlation. 
In contrast, the exact energy given a one-particle basis set can be captured by the FCI method, though FCI is intractable in general as the system size increases.
One class of methods for improving tractability is truncated CI, although these methods suffer from convergence issues towards the FCI limit and lack \textit{size-extensivity}~\cite{helgaker2013molecular}.

An attractive alternative to truncated CI variants is variants based on unitary coupled-cluster (UCC) theory~\cite{bartlett2007coupled,cizek1980coupled,vcivzek1966correlation}, which has been widely adopted in quantum chemistry as a reference baseline for novel quantum algorithms.
In UCC, the ansatz wavefunction is defined as $\ket{\Psi} = e^{\Hat{T}-\Hat{T}^{\dagger}} \ket{\Phi_{\mathrm{HF}}}$, where $\Hat{T}$ is a linear combination of electronic excitations.
In this work, we investigate several variants of the unitary coupled-cluster ansatz, including UCCSD, where the cluster operator $\Hat{T}$ is truncated to include only single and double excitations from occupied to unoccupied spatial orbitals.
Alternatively, the UpCCD approach~\cite{lee2018generalized} only considers double excitations involving both $\alpha$ and $\beta$ spins of the same spatial orbital.
Finally, the UpCCGSD approach uses both singles and double excitations in $\Hat{T}$, as UCCSD, but uses spin-pairing, like in UpCCD, and generalizes the excitation summation over a general set of orbitals.

\subsection{Quantum Teleportation Protocols}\label{subsection:TeleGate}
Within the DQC framework, there are two main primitives for non-local operations: state and gate teleportation~\cite{bennett_teleporting_1993,eisert_optimal_2000}.
These protocols utilize shared Bell pairs, local operations, classical communication, and measurement-conditioned corrections to allow non-local information processing between distinct QPUs. Gate teleportation implements a controlled-unitary gate without requiring direct physical interaction between QPUs. 
In contrast, state teleportation is an identity at the operator level; however, the qubit state information moves between QPUs, which affects the set of available local gates that can be implemented after the teleportation.

\subsection{TeleSABRE}\label{subsection:TeleSABRE}
To run a quantum circuit on a specific quantum architecture, two compilation steps are often used to ensure physical realizability: qubit layout and qubit routing.
This is especially important when a circuit is defined by wires with all-to-all connectivity while a distributed quantum architecture may have limited intra-QPU qubit connectivity and even more limited inter-QPU connectivity. 
The layout step chooses a fixed mapping of circuit wires to physical qubits, while the routing step rewrites the circuit to ensure the connectivity of the distributed quantum system is respected.
The layout and routing algorithm TeleSABRE~\cite{russo2025telesabre} extends the SABRE algorithm~\cite{li_tackling_2019} from a monolithic setting to the distributed quantum computing setting, where QPUs are connected by entanglement generation channels for use in quantum teleportation protocols (gate and state teleportation).
The TeleSABRE algorithm specifically uses a heuristic greedy algorithm with look-ahead to jointly optimize intra-core SWAPs and inter-core teleportation operations, prioritizing less expensive intra-core operations.

\section{Computational Workflow}\label{section:compworkflow}
\label{sec:computational_details}
The UpCCGSD, UpCCD, and UCCSD ans\"{a}tze considered in this work are implemented in Tequila, which generates Qiskit single- and two-qubit-gate circuits and remove any additional classical bits introduced during the conversion process. This circuit compilation step is at the abstract level and not distribution aware.
The choice of circuit parameters does not affect our resource estimation, since we assume our implementable gate set includes arbitrary-angle rotations. For each ansatz and hydrogen-chain system size, we consider two Fermion-to-qubit mappings: Jordan-Wigner (JW)~\cite{jordan1928paulische} and Bravyi-Kitaev (BK)~\cite{bravyi2002fermionic}. 

In this work, we use the default TeleSABRE settings, where routing is capped at 100,000 iterations across up to 100 attempts, dynamically initializing qubit layouts via a round-robin approach and Hungarian matching while enforcing minimum core capacity thresholds. 
Candidate routing moves are guided by a weighted 20-gate lookahead heuristic that incentivizes long-range teleportation with heavy bonuses, balanced against minor operational penalties to avoid local minima.
Finally, multi-core constraints are managed through edge weighting, full-core penalties, passing-core teleportation, and deadlock resolution mechanisms capped at 1,000 steps.

We note that memory constraints limit our results during resource estimation. 
UCCSD becomes intractable to estimate beyond H$_{18}$ under both the naive partitioning strategy and TeleSABRE.
UpCCGSD remains tractable for the naive partitioning strategy up to H$_{50}$, but becomes intractable for TeleSABRE specifically beyond H$_{26}$, owing to the rapid growth of the routing algorithm's search space for its denser two-qubit gate structure. 
No data points are reported for these ansatz-chain-length combinations beyond their respective limits.
All data from this study are available in the project GitHub repository: \url{https://github.com/MSRG/dqc-chem-bench}.

\section{Results and Discussion}\label{section:resultsanddiscussion}
In this section, we examine the suitability of the UpCCD, UpCCGSD, and UCCSD ans\"{a}tze for distribution across hydrogen chains, using the JW and BK encodings described in Section~\ref{sec:computational_details}.
We perform resource analysis using two strategies, where the first strategy partitions each quantum circuit into two even halves and four even quarters, treating the two-qubit gates that cross a partition boundary as candidate teleported gates, and the second strategy applies TeleSABRE, as described in Section~\ref{subsection:TeleSABRE}.
We report these results below, focusing on how teleportation overhead scales with chain length and how the two partitioning strategies compare against each other and against the circuit-cutting baseline.

\begin{figure}[htpb!]
    \centering

    \includegraphics[width=\linewidth]{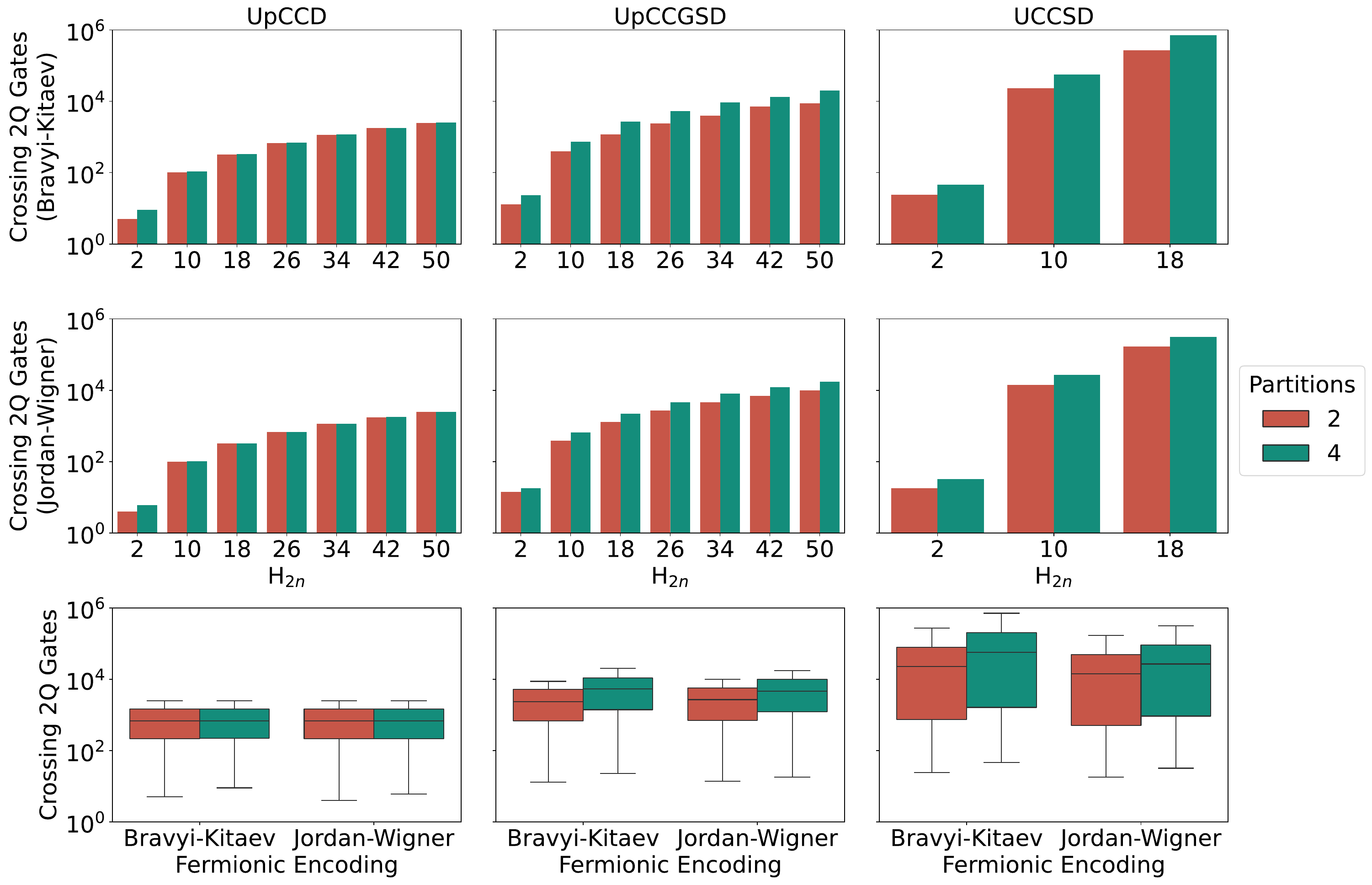}
    \caption{Resource estimations for the three ans\"{a}tze, two fermionic encodings, and two naive partitionings examined. The first two rows show the number of two-qubit gates ($y$-axis) that cross the partitions (hue) as a function of system size ($x$-axis). The third row shows boxplots of the crossing two-qubit gates over all system sizes ($y$-axis) for each fermionic encoding ($x$-axis) and partitioning (hue), where the whiskers denote the minimum and maximum values.}
    \label{fig:resource_comparison}
\end{figure}

Fig.~\ref{fig:resource_comparison} shows the number of two-qubit gates that cross the halved and quartered crossing the naive partitioning strategy, for our three ans\"{a}tze under both the JW and BK encodings, as a function of hydrogen chain length (H$_{2n}$).
Overall, UpCCD requires the least amount of Bell pairs, ranging from 4 (H$_{2}$; JW) to 2508 (H$_{50}$; BK).
UCCSD exhibits the steepest increase in the number of gate crosses with two partitions requiring fewer Bell pairs, while UpCCGSD offers a middle ground between UpCCD and UCCSD.
These trends are summarized further in Fig.~\ref{fig:resource_comparison}, where the UpCCD ansatz offers a consistent number of Bell pairs regardless of the number of partitions or the choice of Fermionic encoding. UpCCGSD shows the same pattern, albeit requiring more Bell pairs, unlike UCCSD, where the BK encoding requires more Bell pairs than the JW encoding.

\begin{figure}[htpb!]
    \centering
    \includegraphics[width=\linewidth]{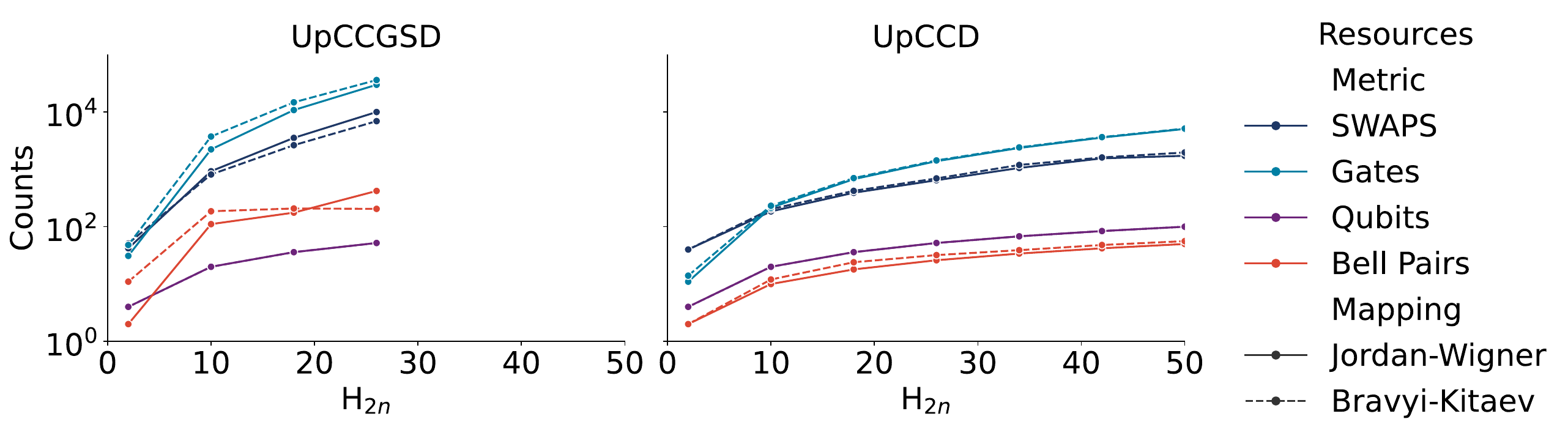}
    \caption{Resource estimations of TeleSABRE for the UpCCGSD (left) and
    UpCCD (right) ans\"{a}tze as a function of hydrogen chain length
    ($\mathrm{H}_{2n}$, $x$-axis). Each panel reports SWAP count, total
    gate count, qubit count, and required Bell pairs (color) on a
    log-scale $y$-axis (Counts) for both the JW (solid lines)
    and BK (dashed lines) encodings. 
    }
    \label{fig:TeleSABREResourceEstimations}
\end{figure}

Fig.~\ref{fig:TeleSABREResourceEstimations} presents the resources needed for the UpCCGSD and UpCCD ans\"{a}tze under the JW and BK encodings as a function of hydrogen chain length ($\mathrm{H}_{2n}$) when routing circuits using TeleSABRE.
Across both ans\"{a}tze, qubit and Bell-pair counts grow much more slowly than SWAP and gate counts, and the encodings track closely with one another throughout, indicating that TeleSABRE's routing and layout optimization largely absorbs the encoding-dependent differences observed under naive partitioning. 
For UpCCD, all four metrics scale similarly across the full range considered. 
For UpCCGSD, which was evaluated up to $\mathrm{H}_{26}$ owing to memory constraints, exhibits substantially larger absolute resource requirements at higher chain lengths. 
This Bell-pair result for UpCCGSD is notably non-monotonic in its encoding dependence: BK requires more Bell pairs than JW at H$_{10}$, a comparable number at H$_{18}$, and less at H$_{26}$; in contrast, with the naive partitioning strategy, BK consistently requires more crossing gates than JW. 
This reversal suggests that TeleSABRE's layout and routing optimization can substantially reshape, and even invert, the encoding-dependent resource trends observed under naive partitioning. 
While the number of two-qubit gates that cross the partitions provides an upper bound for the required number of gate teleportations, these results confirm that routing-aware methods such as TeleSABRE are necessary to obtain realistic, substantially lower resource estimates for distributed execution of these ans\"{a}tze.

\section{Outlook and Conclusions}\label{section:conclusions}
In this work, we systematically observe how the configuration of the ansatz, fermion-to-qubit mapping, and qubit ordering affects overhead and resource scaling for distributed quantum circuits. Among all tested ans\"{a}tze, UpCCD offered the cleanest scaling for all qubit mapping and ordering schemes, while UCCSD demanded the largest resource budget. 
Comparing our two resource-estimation strategies, we find that TeleSABRE consistently reduces the required Bell-pair budget by more than an order of magnitude compared with naive orderings, highlighting the importance of circuit transformations in reducing nonlocality costs.

The above comparison among the tested ans\"{a}tze is based solely on hydrogen chain simulations; generalization beyond this scope is left for future work, which will extend the analysis to additional ans\"{a}tze, including LUCJ and SPA+GS, under TeleSABRE, to larger active spaces, and to more varied molecules.
We also plan to investigate decisions between circuit cutting, teleportation operations, and SWAPs under varying noise conditions.  
Since prior benchmarking of these ans\"{a}tze indicates that the most easily distributed ansatz is not necessarily the most accurate, we plan to directly compare the accuracy-per-resource trade-offs of electronic structure theory circuits, motivating a joint optimization over both distributability and chemical accuracy in future ansatz selection for distributed quantum chemistry.

\begingroup
\small  %
\bibliographystyle{IEEEtran}
\bibliography{quantum_week}
\endgroup

\end{document}